\documentclass[conference,letterpaper]{IEEEtran}
\IEEEoverridecommandlockouts
\usepackage{subcaption}
\usepackage{microtype}
\usepackage{cite}
\usepackage{amsmath,amssymb,amsfonts,amsthm}
\usepackage{mathtools}
\usepackage{graphicx}
\usepackage{textcomp}
\usepackage{xcolor}
\usepackage{array,booktabs,multirow}
\usepackage{siunitx}
\usepackage{algorithm,algpseudocode}
\usepackage[bookmarks=false,bookmarksnumbered=false,hypertexnames=false,hidelinks]{hyperref}
\usepackage{tikz}
\usepackage{pgfplots}
\pgfplotsset{compat=1.14}
\usetikzlibrary{arrows.meta,positioning,fit,shapes.geometric,shapes.misc, backgrounds,calc}
\usepackage{enumitem}

\newcommand{\Sset}{\mathcal{S}}
\newcommand{\Nset}{\mathcal{N}}
\newcommand{\Fset}{\mathcal{F}}
\newcommand{\Uset}{\mathcal{U}}
\def\BibTeX{{\rm B\kern-.05em{\sc i\kern-.025em b}\kern-.08em T\kern-.1667em\lower.7ex\hbox{E}\kern-.125emX}}

\begin{document}

\title{Cross-Slice Co-Location Risk-Aware SFC Provisioning in Multi-Slice LEO Satellite Networks}

\author{Mohammed Mahyoub, Wael Jaafar, Sami Muhaidat, and Halim Yanikomeroglu}

\maketitle
\setlength{\abovedisplayskip}{4pt}
\setlength{\belowdisplayskip}{4pt}
\setlength{\abovedisplayshortskip}{2pt}
\setlength{\belowdisplayshortskip}{2pt}
\begin{abstract}
We address cross-slice co-location risk in multi-slice low Earth orbit (LEO) satellite edge networks, where virtual network functions (VNFs) from different network slices sharing the same satellite instance create a cross-slice security exposure channel. We formulate a risk-aware service function chain (SFC) placement problem as a mixed-integer linear program (MILP) over a dynamically evolving LEO satellite constellation, jointly optimizing cross-slice co-location risk, CPU resource consumption, and VNF migration stability under satellite capacity, inter-satellite link (ISL) capacity, visibility, and end-to-end (E2E) delay constraints. The risk model employs a multiplicative co-location formulation, inspired by the risk assessment principles from ISO/NIST frameworks, with exact and coarse (slice-level) formulations that analytically establish bounds on the co-location exposure. To solve this problem, we propose a three-stage hybrid optimizer combining time epoch preprocessing, simulated annealing-based warm-start, and branch-and-bound refinement. 
Experimental evaluation demonstrates a $40\%$ reduction in co-location risk and an $80\%$ reduction in avoidable VNF migrations relative to the greedy baseline at negligible CPU overhead, and a $23\times$ warm-start speedup from $256$\,s cold-start to $11$\,s per epoch, confirming real-time viability from the second epoch.

\end{abstract}


\vspace{-6pt}
\section{Introduction}\label{sec:intro}

The rapid growth of global connectivity demands and emerging applications, e.g., autonomous systems, maritime communications, and large-scale Internet-of-Things (IoT), has accelerated the development of Non-Terrestrial Networks (NTNs) as a key pillar of 6G \cite{Abdelsadek2023}. Among NTN segments, Low Earth Orbit (LEO) satellite constellations are attracting particular attention and emerging as edge computing platforms for next-generation services, due to their global coverage, low propagation delay, and flexible service provisioning \cite{Inigo2019}.  Enabling multi-slice service provisioning on these platforms requires integration of network function virtualization (NFV) and network slicing, allowing satellites to host VNFs and instantiate dedicated service function chains (SFCs) for multiple isolated slices \cite{Petrosino2023}.  SFC provisioning, which determines where each ordered VNF sequence is deployed and how traffic traverses it, is consequently a central orchestration task \cite{mahyoub_stars_2025}.

SFC provisioning in LEO satellite networks faces many interdependent challenges absent in terrestrial settings, such as scarse on-board computing that limits the number of simultaneously hosted VNFs \cite{Li2022}, and the highly dynamic LEO topology that triggers frequent inter-satellite link (ISL) availability changes, forcing costly VNF migrations across satellites.  When VNFs from different slices share a satellite instance, cross-slice co-location risk arises, whereby a compromised or inference-susceptible shared instance may expose sensitive functions of unrelated slices \cite{mahyoub_survey_2024,Wang2025}.  Jointly addressing these issues requires a risk-aware orchestration framework that existing SFC provisioning do not provide.

SFC provisioning has been extensively studied in NFV-enabled infrastructures~\cite{Satpathy2025}. Mixed-integer linear programming (MILP)-based approaches yield optimal placements under resource and delay constraints~\cite{Chintapalli2024}, while metaheuristics, including genetic algorithms and simulatefd annealing (SA) address scalability~\cite{Noghani2023}. Also, dynamic settings include migration costs for stability~\cite{Afrasiabi2024,mahyoub_stars_2025}. Nevertheless, most works assume trusted infrastructures and omit security risk from the placement objective~\cite{Wang2025}.
Recent studies explored resource allocation and VNF placement in satellite edge computing. For instance, Doan \textit{et al.}~\cite{Doan2023} applied multi-agent reinforcement learnign (MARL) to SFC placement in LEO networks, while Qin~et~al.~\cite{Qin2023} modelled multi-SFC embedding in ultra-dense hybrid LEO-terrestrial 6G systems. 
Alse, Yue \textit{et al.}~\cite{Yue2023wcnc} addressed delay-aware VNF placement in 6G NTN. 
Network slicing orchestration across hybrid satellite-terrestrial segments was studied in \cite{Esmat2023,Kak2022}. Despite these advances, most works treated security as a binary isolation constraint and did not jointly model the interaction between resource efficiency, migration stability and security placement.
Security-aware VNF placement has been studied in terrestrial and edge contexts~\cite{Petrosino2022,AbdulGhaffar2024, mahyoub_slicing_2025}, where strict isolation rules reduce co-location attack surface but limit resource allocation efficiency. In LEO systems, Ahmad \textit{et. al} \cite{Ahmad2022} surveyed satellite-terrestrial security challenges, and Yue \textit{et. al} \cite{Yue2023} analyzed LEO-specific security risks, neither addressing risk-aware SFC placement. 

To the best of our knowledge, no prior work jointly modeled cross-slice security risk, resource utilization, and migration stability for SFC provisioning in LEO satellite networks. Hence, the main contributions of this paper are as follows:
 \begin{itemize}[noitemsep,topsep=2pt]
    \item  We formulate a multi-objective MILP for SFC provisioning in LEO networks combining cross-slice co-location risk, resource consumption, and migration stability.
    \item We propose a risk modelling framework with exact and coarse-grained formulation, and analytically establish bounds that quantify the approximation gap while significantly reducing computational complexity.
    \item We introduce a novel migration-aware stabilization mechanism that distinguishes unavoidable topology-induced changes from avoidable ones.
    \item We develop a hybrid optimization framework combining SA warm-starting and branch-and-bound refinement for efficient security-aware SFC provisioning.
\end{itemize}

\section{System Model and Problem Formulation}\label{sec:model}

\subsection{Network and Service Model}
We consider a multi-slice LEO satellite network observed at discrete time epochs $t$. At each time epoch, the network is represented as a directed graph $G_t=(\Sset,E_t)$, where $\Sset$ is the set of satellites and $E_t$ the set of active ISLs. 
Each satellite $s \in \mathcal{S}$ is characterized by a finite CPU capacity $\textit{Cap}^{\mathrm{cpu}}_s$ and an inter-satellite link (ISL) neighborhood $\mathcal{H}_s=\{s\}\cup\{\hat{s}:(s,\hat{s})\in E_t\}$, 
where $|\mathcal{H}_s| = 4$, comprising two intra-plane links to the fore and aft satellites within the same orbital plane, and two inter-plane cross-links to the nearest satellites in the adjacent orbital planes \cite{Qin2023}.

A user $u$ from slice $n$ can connect to satellite $s$ only if its elevation angle $\theta^t_{n,u,s}$ satisfies $\theta^t_{n,u,s}\ge\theta_{\min}$ where $\theta_{\min}$ is the minimum elevation angle. The set of visible satellites for user $u$ in slice $n$ at epoch $t$ is $\mathcal{V}^t_{n,u}=\{s:\theta^t_{n,u,s}\ge\theta_{\min}\}$. The corresponding access delay $d^{\mathrm{acc}}_{n,u,s}$ is computed using standard slant-range geometry \cite{mahyoub2026_visibility}. 

Let $\mathcal{N}$ be the set of slices, $\mathcal{U}_n$ the users of slice $n$, $\Fset$ the set of VNF types, and $\mathcal{I}$ is the set of instances for any function. Each slice $n$ requires an ordered SFC $F_n=(f^n_1,\ldots,f^n_{L_n})$ where $f_\ell^n$ is the ordered VNF $\ell$ in the SFC $F_n$ and $L_n$ is the chain length \cite{Wang2025}. 
Each VNF instance $(f,i,s)$ is characterized by activation cost $b^{\mathrm{cpu}}_{f,i,s}$, per-user processing cost $a^{\mathrm{cpu}}_{f,i,s}(n,u)$, and processing delay $\tau_{f,i,s}$. Finally, each user has an end-to-end (E2E) delay budget $\bar{T}_{n,u}$.

To ensure tractability, we adopt the following assumptions: i) Processing delays are deterministic, ii) CPU consumption scales linearly with the number of assigned users, iii) Traffic demand is static in each epoch, and iv) Queueing and congestion are not explicitly modelled. These assumptions are valid for low traffic load scenarios, where satellite utilization remains well below $20\%$, making queueing delays negligible.

\subsection{Decision Variables}
Based on the above network and service model, several decision variables are defined as follows:
\begin{itemize}
\item $\beta_{n,u,\ell}^{i,s} \in \{0,1\}$ equals $1$ if user $u \in \mathcal{U}_n$ of slice $n \in \mathcal{N}$ executes function $f^n_\ell$ in its SFC on instance $i \in \mathcal{I}$ hosted on satellite $s \in \mathcal{S}$, and is $0$ otherwise. Where needed, we write $\beta^{i,s}_{n,u,f}$ with $f = f_\ell^n$ to index by VNF type.
\item $\gamma_{f,i}^s \in \{0,1\}$ equals 1 if an instance $i$ of function type $f \in \mathcal{F}$ is activated on satellite $s \in \mathcal{S}$, and $0$ otherwise.
\item $\zeta_{n,u,\ell}^{s,\hat{s}} \in \{0,1\}$  equals 1 if the traffic of user $u \in \mathcal{U}_n$ is routed between function positions $\ell$ and $\ell+1$ via the link from satellite $s$ to satellite $\hat{s} \in \mathcal{H}_s$  and 0 otherwise.
\item $\mu_{n,u,\ell} \in [0,1]$ indicates whether an avoidable migration occurs for user $u \in \mathcal{U}_n$ at function position $\ell$. It is $1$ if the previous placement was feasible but not retained, and $0$ otherwise.
\end{itemize}
\subsection{Constraints}

To enforce feasibility, the following constraints should be satisfied.
First, each user-function pair is assigned exactly once, i.e., 
\begin{equation}\sum_{i,s}\beta_{n,u,\ell}^{i,s}=1, \quad\forall n \in \mathcal{N}, u \in \mathcal{U}_n, \ell=1,\ldots,L_n.
  \label{eq:unique}
\end{equation}

Second, assignments are valid only if the instance is active, i.e., $\beta_{n,u,\ell}^{i,s}\le\gamma_{f_\ell^n,i}^s$, thus ensuring the activation overhead $b^{\mathrm{cpu}}_{f,i,s}$ is charged whenever any user is assigned to that instance. Without it, the solver could assign users to inactive instances, thus avoiding overhead but yielding infeasible solutions. 

Then, the total CPU usage per satellite, from both instance activation overheads and per-user processing demands, must not exceed its capacity $\mathit{Cap}^{\mathrm{cpu}}_s$, i.e.,
\begin{equation}
\sum_{f,i}b^{\mathrm{cpu}}_{f,i,s}~\gamma_{f,i}^s
+\!\sum_{n,u,\ell,i}a^{\mathrm{cpu}}_{f_\ell^n,i,s}(n,u)~\beta_{n,u,\ell}^{i,s}
  \le \mathit{Cap}^{\mathrm{cpu}}_s,\;\forall s \in \mathcal{S}.
  \label{eq:cap}
\end{equation}

For each consecutive VNF pair $(\ell,\ell{+}1)$, the routing indicator $\zeta_{n,u,\ell}^{s,\hat{s}}\in[0,1]$ equals 1 iff user $(n,u)$ traverses ISL $(s,\hat{s})$ between those two VNF positions.  It is coupled to the placement variables via McCormick linearization:
\begin{align}
\zeta_{n,u,\ell}^{s,\hat{s}}&\ge
  \textstyle\sum_{i}\beta_{n,u,\ell}^{i,s}
  +\sum_{i}\beta_{n,u,\ell+1}^{i,\hat{s}}-1,
\label{eq:cons_lb}\\
\zeta_{n,u,\ell}^{s,\hat{s}}&\le
  \textstyle\sum_{i}\beta_{n,u,\ell}^{i,s},
\label{eq:cons1}\\
\zeta_{n,u,\ell}^{s,\hat{s}}&\le
  \textstyle\sum_{i}\beta_{n,u,\ell+1}^{i,\hat{s}},
\label{eq:cons2}
\end{align}
$\forall\,n,\,u\in\mathcal{U}_n,\,\ell\in\{1,\ldots,L_n\},\,(s,\hat{s})\in E_t$, where~\eqref{eq:cons1} and \eqref{eq:cons2} prevent routing on a link unless both endpoints host the respective functions. When two consecutive VNFs are co-located on the same satellite, no ISL is used, and all $\zeta$ are zero, correctly contributing neither ISL delay to~\eqref{eq:delay} nor flow to~\eqref{eq:isl_cap}. Each consecutive VNF pair $(\ell, \ell+1)$ is assigned a single routing variable $\zeta^{s,\hat{s}}_{n,u,\ell}$, implicitly assuming that inter-VNF traffic traverses a single ISL hop. Consequently, the E2E delay must not exceed the user's delay budget $\bar{T}_{n,u}$, i.e.,
\vspace{-.2cm}
\begin{multline}
\underbrace{\sum_{i=1}^{I}\sum_{s\in\mathcal{V}^t_{n,u}}
d^{\mathrm{acc}}_{n,u,s}\,~\beta_{n,u,1}^{i,s}}_{\text{access delay}}
  \;+\;
  \underbrace{\sum_{\ell=1}^{L_n}\sum_{i=1}^{I}\sum_{s\in\Sset}
\tau_{f_\ell^n,i,s}\,~\beta_{n,u,\ell}^{i,s}}_{\text{processing delays}}\\
  +\;
  \underbrace{\sum_{\ell=1}^{L_n-1}\sum_{s\in\Sset}
    \sum_{\hat{s}\in\mathcal{H}_s}
\delta_{s,\hat{s}}\,~\zeta_{n,u,\ell}^{s,\hat{s}}}_{\text{ISL propagation}}
  \;\le\;\bar{T}_{n,u},
  \quad\forall\,n \in \mathcal{N},\,u\in\mathcal{U}_n.
  \label{eq:delay}
\end{multline}
Moreover, the aggregated flow per ISL link is bounded by the link capacity $C^{\mathrm{ISL}}$, such that
\begin{equation}  \sum_{n\in\mathcal{N}}\sum_{u\in\mathcal{U}_n}
    \sum_{\ell=1}^{L_n}\zeta_{n,u,\ell}^{s,\hat{s}}
  \le C^{\mathrm{ISL}},
  \quad\forall\,(s,\hat{s})\in E_t,
  \label{eq:isl_cap}
\end{equation}
while the ingress VNFs (i.e. VNFs at position $\ell=1$) must be placed on visible satellites, i.e., 
\begin{equation} \sum_{i \in \mathcal{I}}\beta_{n,u,1}^{i,s}=0, \quad\forall\,n \in \mathcal{N},\,u\in\mathcal{U}_n,\;s\notin\mathcal{V}^t_{n,u}.
  \label{eq:vis}
\end{equation}

\subsection{Cross-Slice Co-location Risk Model}
In multi-slice LEO satellite networks, VNFs from different slices may be instantiated on shared satellite resources. While such co-location improves resource efficiency, it may introduce cross-slice security exposure, where vulnerabilities in one slice can affect others through shared infrastructure ~\cite{mahyoub_slicing_2025,Ahmad2022}. This work models such exposure as a relative co-location risk metric, intended to guide placement decisions.

We model cross-slice co-location risk as a function of three key factors: i) Function sensitivity $R[f]$ that captures the security criticality of a VNF type (e.g., encryption or intrusion detection functions are more sensitive than traffic monitoring), ii) Slice criticality $C[n]$ that reflects the importance or impact level of a slice (e.g., mission-critical vs. best-effort services), and iii) Isolation policy coefficient $\Phi[n,n']$ that represents the degree of enforced isolation between slices, where $\Phi=0$ indicates strict isolation and $\Phi=1$ indicates no isolation guarantees. 
These factors are combined as
$w^{\mathrm{risk}}_{n,n',f}=R[f]\cdot\Phi[n,n']\cdot C[n]\cdot C[n']$, ensuring that risk is nullified when any factor is zero (e.g., strict isolation or non-sensitive functions), and increases proportionally when risk factors are present.

The exact risk model captures user-level co-location, where exposure arises when users from different slices share the same VNF instance.
Let binary variable $p_{n,u,n^{\prime},u^{\prime},f}^{i,s}$ indicate whether users $u\in\Uset_n$ and $u'\in\Uset_{n'}$ are both assigned to instance $(f,i,s)$, $\forall n \neq n'$. Consequently, this assignment should respect the following constraints:
\begin{align}
p_{n,u,n^{\prime},u^{\prime},f}^{i,s}&\le\beta_{n,u,f}^{i,s},\label{eq:and1}\\
p_{n,u,n^{\prime},u^{\prime},f}^{i,s}&\le\beta_{n',u',f}^{i,s},\label{eq:and2}\\ p_{n,u,n^{\prime},u^{\prime},f}^{i,s}&\ge\beta_{n,u,f}^{i,s}+\beta_{n',u',f}^{i,s}-1,
\label{eq:and3}
\end{align}
and the exact co-location risk is defined as
\begin{equation}
\mathit{Risk}^{\mathrm{ex}}
=\!\sum_{n<n'}\sum_{f\in F_n\cap F_{n'}}\sum_{i,s}
w^{\mathrm{risk}}_{n,n',f}
\!\!\sum_{u\in\Uset_n,\,u'\in\Uset_{n'}}
\!\!p_{n,u,n^{\prime},u^{\prime},f}^{i,s}.
\label{eq:risk_exact}
\end{equation}
This formulation provides a fine-grained representation of co-location exposure but introduces  $\mathcal{O}({N}|^2\,{U}|^2\,{F}\,{I}\,{S})$ auxiliary binary variables, where $N=|\Nset|$, $U=\max_n|\Uset_n|$, $F=|\Fset|$, $I$ is the maximum number of instances per function type per satellite, and $S=|\Sset|$, which may limit scalability.

To improve scalability, we introduce a slice-level approximation. Speciifcally, let  $z_{n,f}^{i,s}\in\{0,1\}$ indicate whether any user of slice $n$ is assigned to instance $(f,i,s)$. Then, this binary variable is linked to $\beta$ through
\begin{align}
z_{n,f}^{i,s}&\ge\beta_{n,u,f}^{i,s},\quad\forall\,u\in\Uset_n,
\label{eq:z1}\\
\text{and } z_{n,f}^{i,s}&\le\!\sum_{u\in\Uset_n}\beta_{n,u,f}^{i,s}.
\label{eq:z2}
\end{align}
Also, let  $y_{n,n',f}^{i,s}\in\{0,1\}$ indicate whether slices $n$ and $n'$ co-locate on instance $i$. Then, their relation is defined via a logical AND of $z_{n,f}^{i,s}$ and $z_{n',f}^{i,s}$ as follows:
\begin{align}
y_{n,n',f}^{i,s}&\le z_{n,f}^{i,s},\label{eq:y1}\\
y_{n,n',f}^{i,s}&\le z_{n',f}^{i,s},\label{eq:y2}\\
\text{and } y_{n,n',f}^{i,s}&\ge z_{n,f}^{i,s}+z_{n',f}^{i,s}-1.\label{eq:y3}
\end{align}
Hence, the resulting coarse risk bounds approximation are given by
\begin{align}
\mathit{Risk}^{\mathrm{LB}}&=\sum_{n<n',f,i,s}
w^{\mathrm{risk}}_{n,n',f}\,~y_{n,n',f}^{i,s},\label{eq:risk_lb}\\
\text{and } \mathit{Risk}^{\mathrm{UB}}&=\sum_{n<n',f,i,s}
|\Uset_n|~|\Uset_{n'}|\,~w^{\mathrm{risk}}_{n,n',f}\,~y_{n,n',f}^{i,s}.
  \label{eq:risk_ub}
\end{align}
This coarse-grained formulation reduces complexity to  $O(N^2\,F\,I\,S)$, enabling efficient optimization for larger systems.
The coarse formulation provides a bounded approximation of the exact risk, i.e.,
\begin{equation}\mathit{Risk}^{\mathrm{LB}} \le \mathit{Risk}^{\mathrm{ex}} \le \mathit{Risk}^{\mathrm{UB}}.
\end{equation}
Table~\ref{tab:complexity} summarizes the variable-count and solution-quality properties of both formulations.
\begin{table}[h]
\centering\small
\caption{Exact vs.\ Coarse Risk Model Comparison}
\label{tab:complexity}
\setlength{\tabcolsep}{3pt}
\renewcommand{\arraystretch}{1}
\begin{tabular}{@{}lcc@{}}
\toprule
\textbf{Property} & \textbf{Exact~\eqref{eq:risk_exact}}
& \textbf{Coarse~\eqref{eq:risk_lb}}\\
\midrule
Granularity & User-pair & Slice-pair\\
Risk variables          & $O(N^2 U^2 FIS)$ & $O(N^2 FIS)$\\
Variable reduction      & --- & $U^2$\\
Objective optimized     & $\mathit{Risk}^{\mathrm{ex}}$ (exact)
 & $\mathit{Risk}^{\mathrm{LB}}$ (lower bound)\\
Bound guarantee         & Exact value & $\mathit{Risk}^{\mathrm{LB}}\le
                          \mathit{Risk}^{\mathrm{ex}}\le
                          \mathit{Risk}^{\mathrm{UB}}$\\
Primary use case        & small-scale & large-scale\\
\bottomrule
\end{tabular}
\end{table}

\subsection{Migration Stabilization} \label{ssec:migration}

Let $\beta^{\mathrm{prev},i,s}_{n,u,\ell}$ be the prior epoch's
placement (a fixed parameter). The keep-indicator
$k_{n,u,\ell}=\sum_{i,s}\beta^{\mathrm{prev},i,s}_{n,u,\ell}\cdot\beta_{n,u,\ell}^{i,s}$ equals 1 if the assignment is unchanged.  The preprocessing
parameter $\pi_{n,u,\ell}\in\{0,1\}$ indicates whether the prior assignment remains feasible.  A migration is considered avoidable if the previous assignment remains feasible under current visibility and capacity constraints but is not retained by the optimizer. Therefore, the avoidable-migration indicator is
\begin{align}\mu_{n,u,\ell}&\ge\pi_{n,u,\ell}-k_{n,u,\ell},\label{eq:mu1}\\
\mu_{n,u,\ell}&\le\pi_{n,u,\ell}, \text{ and }  
  \mu_{n,u,\ell}\le 1-k_{n,u,\ell}.\label{eq:mu23}
\end{align}
$\mu_{n,u,\ell}\in\{0,1\}$ equals $1$ only when the prior assignment was feasible but was not retained.
Thus, the total migration disruption cost is $\mathit{Mig}=\sum_{n,u,\ell}\mathit{Dis}^{\mathrm{mig}}_{f_\ell^n} \,~\mu_{n,u,\ell}$,
where $\mathit{Dis}^{\mathrm{mig}}_{f_\ell^n}\ge 0$ is the per-type disruption cost, capturing state-transfer overhead and transient service downtime during VNF migration.

\subsection{Optimization Objective}

Let $\mathit{CapUse}$ be the total satellite CPU consumption, consistent with the capacity constraint~\eqref{eq:cap}, i.e.,
\begin{equation}
  \mathit{CapUse}=\!\sum_{s,f,i}b^{\mathrm{cpu}}_{f,i,s}\,\gamma_{f,i}^s  +\!\sum_{n,u,\ell,i,s}a^{\mathrm{cpu}}_{f_\ell^n,i,s}(n,u) \,\beta_{n,u,\ell}^{i,s}.
  \label{eq:capuse}
\end{equation}
The objective is to minimize a normalized and weighted combination of resource consumption, co-location risk, and migration as
\begin{equation}
  \min\;
  \omega_{\mathrm{cap}}\frac{\mathit{CapUse}}{\overline{\mathit{CapUse}}}
  +\omega_{\mathrm{risk}}\frac{\mathit{Risk}}{\overline{\mathit{Risk}}}
  +\omega_{\mathrm{mig}}\frac{\mathit{Mig}}{\overline{\mathit{Mig}}},
  \label{eq:obj}
\end{equation}
where $\omega_{\mathrm{cap}},\omega_{\mathrm{risk}},\omega_{\mathrm{mig}}\ge 0$ with $\omega_{\mathrm{cap}}+\omega_{\mathrm{risk}}+\omega_{\mathrm{mig}}=1$, and the normalization bounds $\overline{\mathit{CapUse}}$, $\overline{\mathit{Risk}}$, and $\overline{\mathit{Mig}}$ are precomputed via analytical upper bounds. The weights allow operators to express policy preferences. For instance, a security-sensitive operator sets a large $\omega_{\mathrm{risk}}$, while an efficiency-driven operator prioritizes $\omega_{\mathrm{cap}}$. The problem is NP-hard, as it contains bin-packing \cite{Garey1979}. Adding cross-slice risk and migration objectives does not alter the complexity class but substantially increases the model size, motivating the following scalability strategies.

\subsection{Scalable Hybrid Optimization}\label{subsec:scalability}

\begin{figure}[!t]
    \centering
    \begin{tikzpicture}[
      box/.style={rectangle,draw=black!62,rounded corners=4pt,
                  fill=blue!9,thick,align=center,minimum width=2.55cm,
                  minimum height=0.74cm,font=\small},
      obox/.style={box,fill=orange!18,draw=orange!68!black},
      gbox/.style={box,fill=green!12,draw=green!52!black},
      arr/.style={-Stealth,thick,black!62},
      darr/.style={-Stealth,thick,dashed,gray!52},
      lbl/.style={font=\scriptsize,align=center}
    ]
    \node[box](snap) at(0,2.9){Network Snapshot\\[-2pt]\scriptsize epoch $t$,
      $G_t$, $\mathcal{V}^t_{n,u}$};
    \node[box](prep) at(3.25,2.9){Pre-processing\\[-2pt]\scriptsize
      $\pi_{n,u,\ell}$, bounds, norms};
    \node[obox](sa)  at(0,1.22){Simulated Annealing\\[-2pt]\scriptsize
      fast feasible $\beta^{SA}$};
    \node[obox](ws)  at(3.25,1.22){MILP Warm-Start\\[-2pt]\scriptsize
      incumbent: $\beta^{SA}$/$\beta^{\mathrm{prev}}$};
    \node[gbox](milp)at(1.62,0.0){MILP Refinement\\[-2pt]\scriptsize
      branch-and-bound + cuts};
    \node[box](dep)  at(1.62,-1.12){Deployment\\[-2pt]\scriptsize
      $\beta,\gamma,\zeta,\mu$};
    \draw[arr](snap)--(prep); \draw[arr](snap)--(sa);
    \draw[arr](prep)--(ws); \draw[arr](sa)--(ws);
    \draw[arr](ws)--(milp);
    \draw[arr](sa.south) to[out=-72,in=172](milp.west);
    \draw[arr](milp)--(dep);
    \draw[darr](dep.south) to[out=-90,in=-90]
      node[above,lbl,yshift=-1pt]{store $\beta^{\mathrm{prev}}$}
      +(-3.85,0) to[out=90,in=180](snap.west);
    \begin{scope}[on background layer]
      \node[draw=blue!32,fill=blue!3,dashed,rounded corners=5pt,
            fit=(sa)(ws)(milp),inner sep=5pt,
            label={[font=\scriptsize,blue!52]above:Hybrid Optimization Engine}]{};
    \end{scope}
    \end{tikzpicture}
    \caption{Proposed 3-stage hybrid optimization workflow. Preprocessing computes the visibility set $\mathcal{V}^t_{n,u}$, feasibility indicators $\pi_{n,u,\ell}$, and objective normalization bounds. SA generates a fast feasible incumbent $\beta^{SA}$, and the MILP refines it via branch-and-bound. At each epoch, $\beta^{\mathrm{prev}}$ is updated for the subsequent epoch's warm-start.}
\label{fig:workflow}
\end{figure}
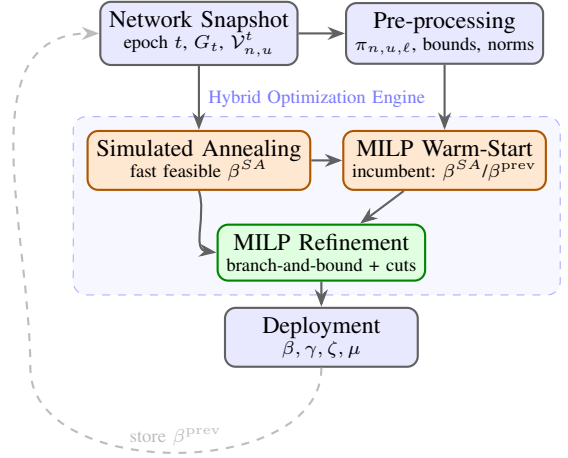

The 3-stage hybrid optimizer shown in Fig. \ref{fig:workflow} handles the NP-hardness of the MILP within practical 60-second epoch budgets as follows:

\subsubsection{Stage 1 - Preprocessing}
At each epoch, visibility sets $\mathcal{V}^t_{n,u}$, feasibility flags $\pi_{n,u,\ell}$, and normalization bounds
$(\overline{\mathit{CapUse}},\overline{\mathit{Risk}},\overline{\mathit{Mig}})$ are computed via upper bounds.

\subsubsection{Stage 2 - SA warm-start}
The SA begins from a greedy feasible placement constructed as follows. For each user $u \in \mathcal{U}_n$ and function chain position $\ell$, the ingress VNF is assigned to the nearest satellite in the visibility set $\mathcal{V}^t_{n,u}$ that has sufficient residual CPU capacity. Each subsequent VNF is assigned to the satellite with the smallest hop count from the ingress satellite that again satisfies the CPU constraint. At each iteration $k \in \{1,\dots,K\}$, a single user-satellite pair is chosen uniformly at random, and its current satellite assignment is proposed to change from $s$ to a uniformly sampled candidate $s' \in \mathcal{V}^t_{n,u}$.
The move is accepted via the Metropolis criterion, i.e.,
\begin{equation}
  P(\text{accept})=
  \begin{cases}1 & \Delta\le 0,\\
  \exp(-\Delta/T_k) & \text{otherwise,}\end{cases}
  \label{eq:metropolis}
\end{equation}
where $\Delta$  is the change in the normalized weighted objective~\eqref{eq:obj} induced by the proposed reassignment, and $T_k = T_0\!\left(T_{\mathrm{end}}/T_0\right)^{k/K}$ is the temperature at iteration $k$ under a geometric cooling schedule, with $T_0$ and $T_{\mathrm{end}}$ denoting the initial and final temperatures, respectively, and $K$ the total number of SA iterations. To avoid a full objective recompute at every iteration, the objective change $\Delta$ is evaluated incrementally. Relocating $(n,u,\ell)$ from $s$ to $s'$ affects only three localized quantities: (i) the CPU loads on exactly two satellites, $s$ and $s'$, which are updated by adding or removing the activation and per-user CPU of the moved VNF, (ii) the co-location risk, whose delta is computed in $O(|\mathcal{U}_n|)$ steps by scanning only the users of the same function type already hosted on $s$ and $s'$, and (iii) the E2E delay for user $u$ alone, recomputed along the updated satellite sequence using precomputed shortest-path delays. Moves that violate CPU capacity or the E2E delay budget are rejected without computing $\Delta$.

\subsubsection{Stage 3 - MILP refinement} The SA solution $\beta^{SA}$ warm-starts branch-and-bound with a $0.5\%$ optimality gap.
This tolerance was chosen as the tightest gap achievable within the residual epoch budget across all evaluated epochs. The previous epoch's placement $\beta^{\mathrm{prev}}$ serves as a fallback incumbent when preprocessing confirms its continued feasibility.

\section{Simulation Setup}\label{sec:setup}
 
 The full implementation of this work is publicly accessible to support reproducibility and enable future research\footnote{\href{https://github.com/Mahyub/LEO_SFC_PROVISIONING}{https://github.com/Mahyoub/LEO-SFC-PROVISIONING}}. 
 
 A Walker-Star constellation with $|\Sset|=60$ satellites across $4$ orbital planes ($15$ per plane), at $550$\,km altitude and $53^\circ$ inclination is simulated using MATLAB  Satellite Communications Toolbox. $15$ consecutive $60$-second epochs (total of $15$ minutes) are simulated. At $550$ km altitude, one orbital period is approximately $95$ minutes. The $15$-epoch window captures the most dynamic phase of each pass, satellite rise, peak visibility, and is sufficient to observe multiple ISL topology changes. The minimum elevation angle is set to $10^\circ$  according to 3GPP recommendations \cite{3gpp38811}.

$N=|\mathcal{N}|=5$ network slices are anchored at geographically dispersed ground stations (London, New York, Tokyo, Sydney, Paris), each serving $|\mathcal{U}_n|=10$ users, $\forall n=1,\ldots,5$, as the current evaluation focuses on realistic initial deployments.  $F=|\mathcal{F}|=5$ VNF types are used, such that
$\Fset=\{\text{FW},\text{IDS},\text{ENC},\text{TM},\text{SIEM}\}$. Each SFC has length $L_n\in\{2,3,4\}$, drawn uniformly with repeated types permitted (motivating position-based indexing). Finally, E2E budgets are  $\bar{T}_{n,u}\sim\text{Uniform}[75,150]$\,ms.
Function risk sensitivities $R[f]$ are calibrated against NIST SP 800-53 security impact levels as follows: ENC $\rightarrow R=0.9$ (HIGH: key material exposure), IDS $\rightarrow  R=0.8$ (HIGH: detection evasion), FW $\rightarrow  R=0.6$ (MODERATE: policy bypass), SIEM $\rightarrow  R=0.6$ (MODERATE: log tampering), TM $\rightarrow  R=0.4$ (LOW: passive monitoring). These mappings follow the confidentiality impact ratings from NIST SP 800-53 control families SC (System and Communications Protection) and AU (Audit and Accountability).
Also, slice criticality $C[n]\sim\text{Uniform}[1,3]$, following 3GPP TS~22.261 tiers, and the isolation policy $\Phi[n,n']\sim\text{Uniform}[0,1]$ per ETSI GS~NFV-SEC~026. 

For the SA approach, initial temperature $T_0 = 1$, final temperature $T_{end} = 0.01$, and the iteration limit $K = 50000$. The chosen values minimized the normalized objective while keeping the SA runtime below $2$ sec. Moreover, we set the objective weights as
$\omega=(\omega_{\mathrm{cap}},\omega_{\mathrm{risk}},\omega_{\mathrm{mig}})=(0.3,0.5,0.2)$, reflecting a security-centric deployment scenario. Finally, $3$ baselines are considered, namely B1 (resource-min MILP, $\omega=(1,0,0)$), B2 (stability-aware MILP, $\omega=(0.5,0,0.5)$, no risk term), and B3 (greedy nearest-satellite heuristic).  

\section{Experimental Results}\label{sec:results}
\begin{figure}[!t]
  \centering  \includegraphics[width=0.75\linewidth]{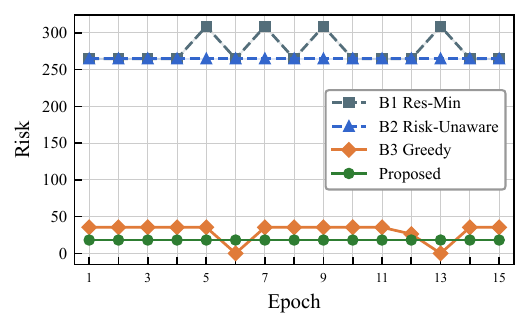}
    \caption{Per-epoch risk over $15$ epochs for each method.}
    \label{fig:risk_epochs}
\end{figure}

Fig.~\ref{fig:risk_epochs} presents the per-epoch evolution of
the risk as a line series over the $15$-epoch simulation window. As shown, the proposed method achieves risk reduction to $18.12$ constant across all 15 epochs, representing a $93.4\%$ reduction versus B1 (risk of $276.6$) and $93.2\%$ decrease versus B2 (risk of $264.9$).  Even though B3 benefits from incidental geographic separation, the proposed method reduces risk by $39.8\%$ compared to it. The constant risk across epochs demonstrates that the risk-minimizing placement pattern is topologically robust under Walker-Star orbital dynamics. However, the elevated risks of B1 and B2 are a direct consequence of their consolidation objectives. Indeed, minimizing CPU activation overhead packs all slices onto minimal satellites, thus maximizing cross-slice instance sharing. Moreover, B1 shows risk spikes (up to $308.9$) at topology-change epochs where re-consolidation yields an even more co-location-intensive configuration. 

\begin{figure}[!t]
    \centering
    \includegraphics[width=0.8\linewidth]{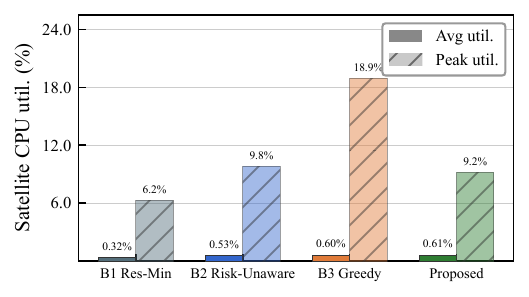}
    \caption{Mean active-satellite CPU utilization (light bars) and mean peak per-satellite CPU utilization (dark bars).}
    \label{fig:peak_vs_avg_util}
\end{figure}

Fig.~\ref{fig:peak_vs_avg_util} provides a grouped bar chart of both average CPU utilization (Avg util.) and peak per-satellite utilization (Peak util.) to reveal hot-spot behaviour.
The proposed method consumes $0.61\%$ total constellation CPU, versus $0.32\%$ for the most efficient baseline (B1). This gap is the quantitative ``price of security''. Indeed, in this case, the $93\%$ risk reduction costs only $0.29\%$ CPU usage.  The risk-motivated geographic spread also provides implicit load balancing. Also, the proposed method's peak per-satellite CPU (at $9.2\%$) is below both B2 and B3's hotspot peaks, as the letter co-locate more VNFs within a smaller number of satellites. 

\begin{figure}[t]
  \centering
  \includegraphics[width=0.75\columnwidth]{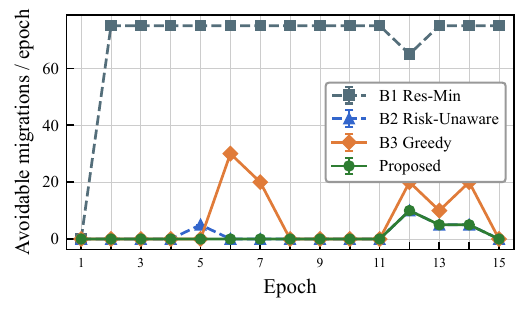}
  \vspace{-.3cm}
  \caption{Avoidable VNF migrations per epoch for each method.}
  \label{fig:migration_epochs}
\end{figure}

Fig.~\ref{fig:migration_epochs} illustrates avoidable VNF migrations per epoch over the $15$-epoch window. B1 incurs an average of $69.3$ migrations per epoch, essentially restarting placement from scratch every $60$\,s, while B3 still accumulates an average of $6.7$ due to its lack of migration awareness. In contrast, the proposed method achieves an average of $1.33$ avoidable migrations per epoch, concentrated at epochs with significant topology changes. Indeed, $12$ of the $15$ epochs show zero unnecessary reassignments. The $80\%$ reduction relative to B3 and $98.1\%$ relative to B1 confirms that the keep-indicator mechanism effectively suppresses unnecessary placements.

\begin{figure}[!t]
  \centering
    \centering
    \includegraphics[width=0.75\linewidth]{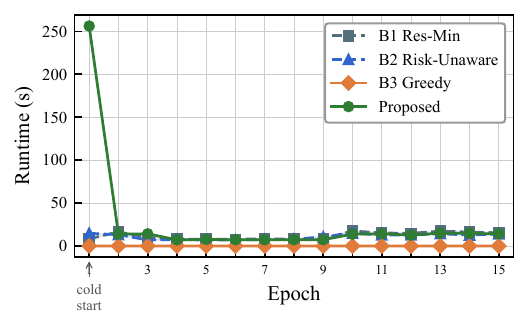}
    \vspace{-0.4cm}
    \caption{Per-epoch solve runtime for each method.}
    \label{fig:runtime_epochs}
\end{figure}

Fig.~\ref{fig:runtime_epochs} presents the per-epoch runtime trajectory to reveal warm-start benefits. The cold-start at the first epoch requires $256$s, exceeding the $60$-second epoch budget. This occurs once at system initialization, before any prior placement $\beta^{prev}$ is available. In operational deployments, this can be handled by (i) pre-computing an initial placement during the preceding handover window using a coarser LP relaxation as an incumbent, or (ii) deferring service activation by one epoch. From epoch $2$ onwards, SA warm-starting ($0.5-1.8$s) reduces the mean MILP runtime to $11.1$\,s, a $23\times$ reduction, confirming practical viability within $60$-second epoch intervals. The proposed method's runtime matches the simpler baselines (B1 at $12.0$\,s and B2 at $10.9$\,s), demonstrating that the additional risk variables impose insignificant material overhead once warm-start benefits are realized. B3 solves in ${\approx}4$\,ms per epoch via its $\mathcal{O}(S\,U\,L)$ polynomial heuristic.

\begin{figure}
    \centering
    \includegraphics[width=0.75\linewidth]{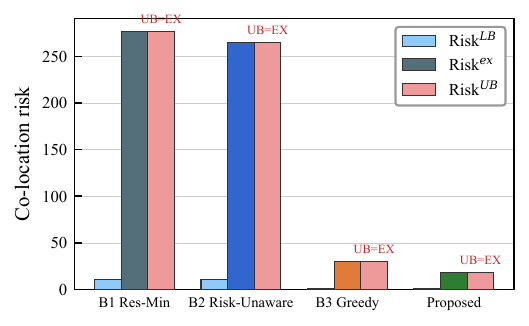}
    \caption{Mean $\mathit{Risk}^{\mathrm{LB}}$, $\mathit{Risk}^{\mathrm{ex}}$, and $\mathit{Risk}^{\mathrm{UB}}$ for each method.}
    \label{fig:bound_tightness}
\end{figure}

Finally, Fig.~\ref{fig:bound_tightness} shows $\mathit{Risk}^{\mathrm{LB}}$, $\mathit{Risk}^{\mathrm{ex}}$, and $\mathit{Risk}^{\mathrm{UB}}$ for each method. 
The coarse upper bound $\mathit{Risk}^{\mathrm{UB}}$ equals
$\mathit{Risk}^{\mathrm{ex}}$ exactly for all methods.  This equality holds because the risk-aware optimizer avoids dense co-location, ensuring at most one user per slice occupies any shared instance, collapsing the upper bound to the exact value. For resource-minimizing baselines, the same equality holds due to the small number of users per slice relative to available instances. Despite the loose lower bound, the coarse model drives placement decisions that achieve near-optimal exact risk, confirming its practical effectiveness as an optimization proxy.

\section{Conclusion and Future Work}
\label{sec:conclusion}


This paper formulated risk-aware SFC placement in multi-slice LEO satellite edge networks as an MILP jointly minimizing cross-slice co-location risk, resource consumption, and migration disruption. A 3-stage hybrid optimizer (epoch preprocessing \& SA warm-start \& branch-and-bound) achieved a $23.0\times$ per-epoch speedup, delivering $40\%$ co-location risk reduction and $80\%$ avoidable-migration reduction versus the greedy baseline at negligible CPU overhead, with sub-$12$-second re-optimization within $60$-second orbital epochs.  
While this work assumes static slices and fixed E2E delay budgets, future research should address dynamic slice arrivals and departures in rapidly evolving 6G networks. 

\bibliographystyle{IEEEtran}
{\footnotesize\setlength{\itemsep}{0pt}
\bibliography{references}}

\end{document}